\documentclass[journal]{IEEEtran}
\usepackage{graphicx}
\begin{document}
\thispagestyle{empty}
\title{Modeling the field of laser welding melt pool by RBFNN}

\author{\authorblockN{Anamarija Bor\v stnik Bra\v ci\v c, Edvard Govekar, Igor Grabec}\\
\authorblockA{Faculty of Mechanical Engineering, University of Ljubljana,\\
A\v sker\v ceva 6, POB 394, SI-1001 Ljubljana, Slovenia\\ 
Email: anamarija.bracic@fs.uni-lj.si}
\thanks{Manuscript received: January 31, 2007}}
\maketitle
\thispagestyle{empty}

\begin{abstract}
Efficient control of a laser welding process requires the reliable
prediction of process behavior. A statistical method of field modeling,
based on normalized RBFNN, can be successfully used to predict 
the spatiotemporal dynamics of surface optical activity in the laser welding process.
In this article we demonstrate how to optimize RBFNN to 
maximize prediction quality. Special attention is paid to the structure of 
sample vectors, which represent the bridge between the field distributions in the past and future.
\end{abstract}


\section{Introduction}
Laser systems are efficiently applied in welding processes\,\cite{govekar1}, 
where a laser beam is used to melt material.  
To maintain high performance in a welding process, efficient control should be established. 

The crucial task in planning the control system is to determine representative variables which can effectively describe the welding process. 
For this purpose, the intensity and spatial distribution of reflected light, surface temperature values or properties of the emitted electron plasma 
are usually chosen. However, characteristic dynamic properties in space and time can also be obtained by recording surface optical activity 
in the heated zone, known as the melt pool\,\cite{govekar2}. 

After choosing the representative variables, extraction of evolution laws from temporal data becomes a crucial problem. To date, this problem 
has been extensively studied in relation to chaotic time series prediction\,\cite{casdagli,abarbanel,kosterlich,kantz,sigert}. The basis  
of these methods is to reconstruct a state-space from a recorded scalar time series by using an embedded technique, and then to estimate 
deterministic dynamic evolution from the reconstructed trajectory using statistical average estimators. We present a generalization of this approach, 
where the modeling of dynamic laws is extended from one dimension (time) to multiple dimensions in spatiotemporal space. This generalization 
requires a new embedding method, which makes feasible a reconstruction of trajectory in the state-space from spatiotemporal data. 
The embedded technique, which was initially developed for time series analysis, can be simply generalized to spatially related data\,\cite{rubin,grabecmandelj,orstavik,parlitz,mandelj} 
and results in a good agreement between predicted and original chaotic fields over short time scales. Since, in a properly reconstructed 
state-space, the modeled dynamics must have similar statistical properties to the actual dynamics, we use a new state-space reconstruction 
method which also considers statistical properties of a field structure. Such reconstruction results in an accurate short-term prediction 
as well as a statistically proper long-term prediction of deterministic chaotic field evolution\,\cite{mandeljclanek}.

In this article, a statistical method of field generators, which is based on 
normalized \emph{radial basis function neural network} (RBFNN), 
is used to model the spatiotemporal dynamics of laser welding melt pool images. 
The stochastic field evolution is modeled 
from sample state vectors 
reconstructed from recorded spatiotemporal data. The field evolution equation is estimated non-parametrically from the samples, using the 
conditional average estimator which determines the governing equation of RBFNN. The goal of this article is to find an optimal dimensionality 
of the neural network, {\em i.e.}, to determine its optimal structure and an adequate number of sampling patterns, which will result 
in the best quality $Q$ of field generator prediction.

Accurate modeling of laser welding images, together with a criterion function specified by the operator of the laser system, provides the 
basis for optimal control of the laser welding process.

\section{Description of RBFNN}
\subsection{Non parametric statistical modeling}
Experimental analysis of process dynamics is based on a representative record of the field ${\bf \varphi}={\bf \varphi} ({\bf s})$, 
where the variable ${\bf s}$ represents space as well as time components ${\bf s}= {\bf s}({\bf r},$t$)$. Most commonly, the spatiotemporal field 
evolution of ${\bf \varphi} ({\bf s})$ is described analytically by a system of nonlinear partial differential equations or 
integrodifferential equations. An analytical form of the model can be estimated from 
the recorded data, based on spatial and temporal derivatives \,\cite{vos98,bar,vos99}. In the case of experimentally 
obtained data, it is difficult to estimate derivatives. Therefore, for a more general approach, a model of field evolution 
should be expressed in terms of recorded data only. 

\begin{figure}
\centering
\includegraphics[width=3.5in]{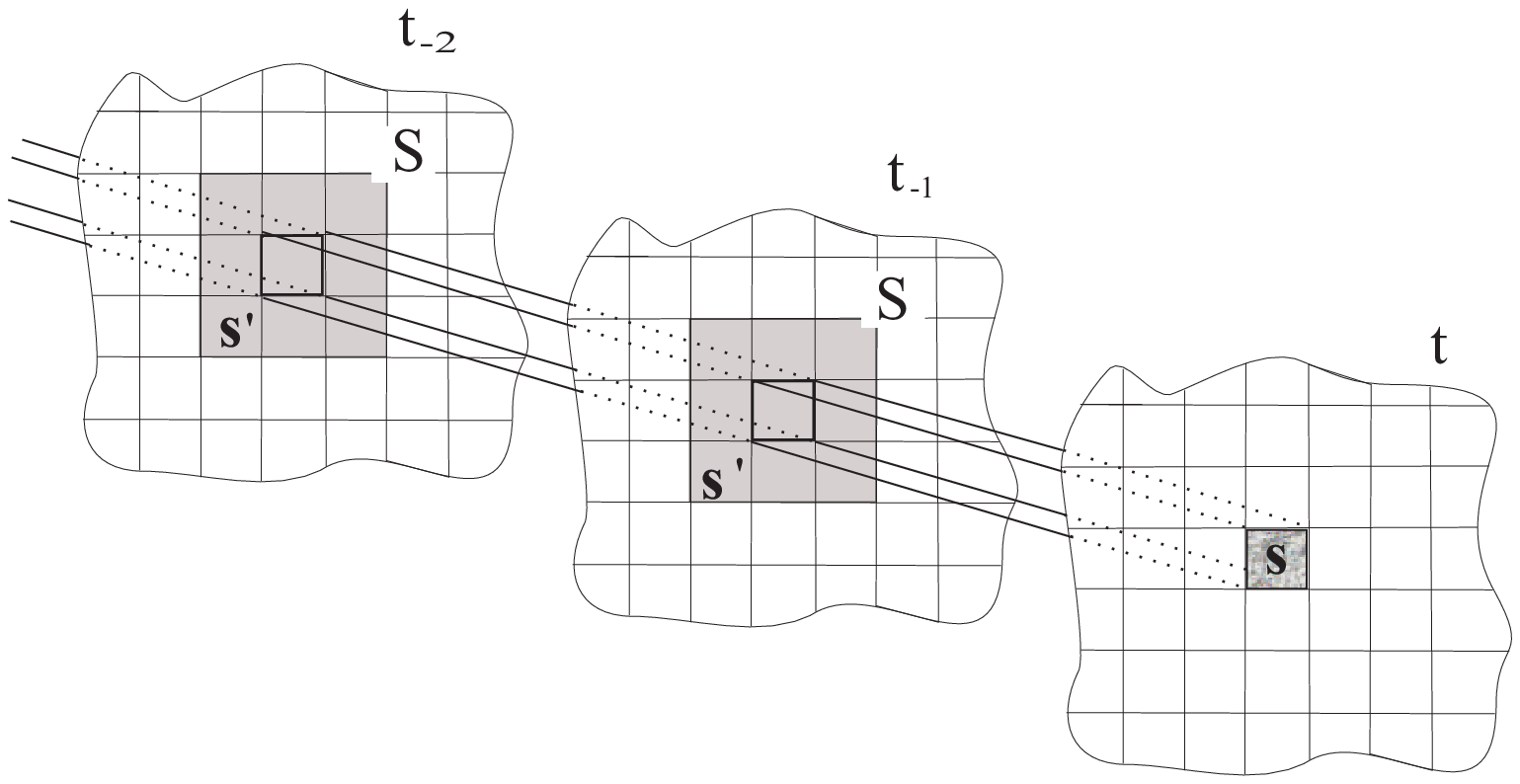}
\caption{Illustration of point ${\bf s}$  and its surroundings ${\bf s}' \in {\cal S}$. 
The future distribution of field ${\bf \varphi}$ (${\bf s}$) is located in  plane $t_{+1}$, while the surrounding points 
${\bf s}' \in {\cal S}$, which represent the past distribution of field, are located in {\em  planes} $t$, $t_{-1}$, $t_{-2}$...}
\label{okolica}
\end{figure}

In our model, field evolution is expressed in terms of data  recorded at equally spaced discrete points in 
space and time. We assume that the dynamics of the field can be described in terms of the generator equation
\begin{equation}
{\bf \varphi} ({\bf s})= {\cal G}\left({\bf \varphi}({\bf s'} \in {\cal S} ({\bf s})),\sigma \right),
\label{e1}
\end{equation}
where ${\bf \varphi}({\bf s'} \in {\cal S})$ represents the {\it past distribution} of the record, while 
${\bf \varphi}({\bf s})$ represents its {\it future distribution}. ${\cal S}$ represents the surroundings of 
point ${\bf s}$. The field generator ${\cal G}$ provides for determination of  the future field distribution 
from its past distribution. $\sigma$ is a model parameter depending on the experimental setup and will be specified in greater detail later. 
An arbitrary point ${\bf s}$ and its surroundings ${\bf s}' \in {\cal S}$ are illustrated in Fig.\,\ref{okolica}.
The source of information for modeling the field generator is a field record containing joint sample pairs 
${\bf \varphi}({\bf s})$ and ${\bf \varphi}({\bf s'}  \in {\cal S})$. These joint sample pairs form a sample vector 
${\bf V}_i({\bf s}) = ({\bf \varphi}_i({\bf s}), {\bf \varphi}_i({\bf s}' \in {\cal S}))$. 
To make further derivation more transparent, the past field distribution ${\bf \varphi}_i({\bf s'})$ and the future 
field distribution ${\bf \varphi}_i({\bf s})$  will be denoted by ${\bf x}_i$ and ${\bf y}_i$, respectively. Hence 
${\bf V}_i({\bf s}) = ({\bf y}_i,  {\bf x}_i)$. 

The samples ${\bf V}_i$ are interpreted as random variables and can therefore be used to express the 
{\it joint probability distribution function} (PDF) by the kernel estimator\,\cite{duda3}
\begin{equation}
f_N ({\bf V})= \frac{1}{N} \sum_{i=1}^{N} \psi({\bf V}-{\bf V}_i,\sigma),
\label{PDF}
\end{equation}
in which $\psi$  denotes an acceptable kernel function such as the Gaussian function $\psi(x-x_i,\sigma) = 1/(\sqrt{2 \pi} \sigma) {\rm exp}(-(x-x_i)^2/2 \sigma)$ 
and $N$ is the number of sample pairs.

Once the samples from the field record have been taken, the question of {\em how to determine the optimal predictor} becomes relevant. 
We consider as an optimal predictor of the future field distribution ${\bf y}$ from a given value ${\bf x}$  
the value ${\bf \hat{y}}$  at which the mean square prediction error is minimal:
\begin{equation}
{\rm E}[ ({\bf y}-{\bf \hat{y}})^2 |x ] = {\rm min}({\bf \hat{y}}).
\label{minE}
\end{equation}
Here ${\rm E}[\,\,]$ denotes averaging over all points in a field record at a given time $t$. The solution of Eq.\,(\ref{minE}) 
yields together with PDF from Eq.\,(\ref{PDF}) {\em the conditional average estimator}
\begin{equation}
{\bf \hat{y}} ({\bf x}) = \frac{\sum_{i=1}^{N} {\bf y}_i \psi ({\bf x} - {\bf x}_i, \sigma)}{\sum_{j=1}^{N}  \psi ({\bf x} - {\bf x}_j, \sigma)} = \sum_{i=1}^{N} {\bf y}_i C_i({\bf x}),
\label{CE}
\end{equation}
where coefficients of the expansion $C_i({\bf x})$ represent basis functions that measure the similarity between the temporary 
vector ${\bf x}$ and 
vector ${\bf x}_i$ from the field record.   
The conditional average estimator described by Eq.\,\ref{CE} represents a radial basis function neural network in which the 
recorded data ${\bf x}_i,{\bf y}_i$ represent the memorized contents of neurons, ${\bf x}$ and ${\bf \hat{y}} ({\bf x})$ are the 
input and the output of the network, while the basis functions $C_i({\bf x})$ correspond to activation functions of neurons. 
Since $\sum_{i=1}^{N} C_i({\bf x}) = 1 $, the conditional average estimator  represents a normalized RBFNN. In this function, the parameter $\sigma$ can be interpreted as the width of receptive fields of neurons.

\subsection{Quality of predictor}

Working towards  optimal modeling of future field distributions requires a quantitative estimation of modeling quality. 
We therefore 
introduce {\em a testing field} ${\bf y}$ and define the  {\em prediction quality} $Q$,  based upon  the difference between the predicted field 
${\bf \hat{y}}$ and the testing field ${\bf y}$ as:
\begin{equation}
Q = 1 - \frac{{\rm E}[({\bf \hat{y}}-{\bf y})^2]}{{\rm E}[({\bf \hat{y}}-{\bf \hat{\bar{y}}})^2]+{\rm E}[({\bf y}-{\bf \bar{y}})^2]}.
\label{Q}
\end{equation}
Here $ {\bf \hat{\bar{y}}}$ and ${\bf \bar{y}}$ stand for the average values of predicted field ${\bf \hat{y}}$ and 
testing field ${\bf y}$, {\em i.e.}, ${\rm E}[{\bf \hat{y}}] = {\bf \hat{\bar{y}}}$ 
and ${\rm E}[{\bf y}] = {\bf \bar{y}}$.
A perfect prediction ${\bf \hat{y}}={\bf y}$ yields $Q=1$, while uncorrelated ${\bf \hat{y}}$ and ${\bf y}$ result in $Q=0$.

\subsection{Prediction of field evolution}

The prediction process consists of three steps: 
\begin{enumerate}
\item {\em Learning}, that corresponds to {\em setting up the basis of joint sample pairs} 
$({\bf \varphi}_i({\bf s}), {\bf \varphi}_i({\bf s}' \in {\cal S}))$ = $({\bf x}_i$, ${\bf y}_i)$ from the field record,
\item {\em predicting the field} ${\bf \hat{y}}$ by using the conditional average estimator from Eq.\,(\ref{CE}), 
\item and, if the testing field exists, {\em comparing predicted field 
with testing field} and 
calculating prediction quality $Q$.
\end{enumerate}
 
In order to achieve the highest quality of prediction for the process, answers to the following crucial questions are needed:  
\begin{itemize}
\item  How to find the surrounding ${\cal S}$ of a given point ${\bf s}$, which gives the best prediction 
of field ${\bf \hat{\varphi}}({\bf s})$ at this point? 
\item  How to determine an optimal number of joint sample pairs $({\bf \varphi}_i({\bf s}), {\bf \varphi}_i({\bf s}' \in {\cal S}))$ = $({\bf x}_i$, ${\bf y}_i)$? 
\end{itemize}
These questions will be addressed in the following chapters.

\section{Time evolution of melt pool}

\begin{figure}
\centering
\includegraphics[width=3in]{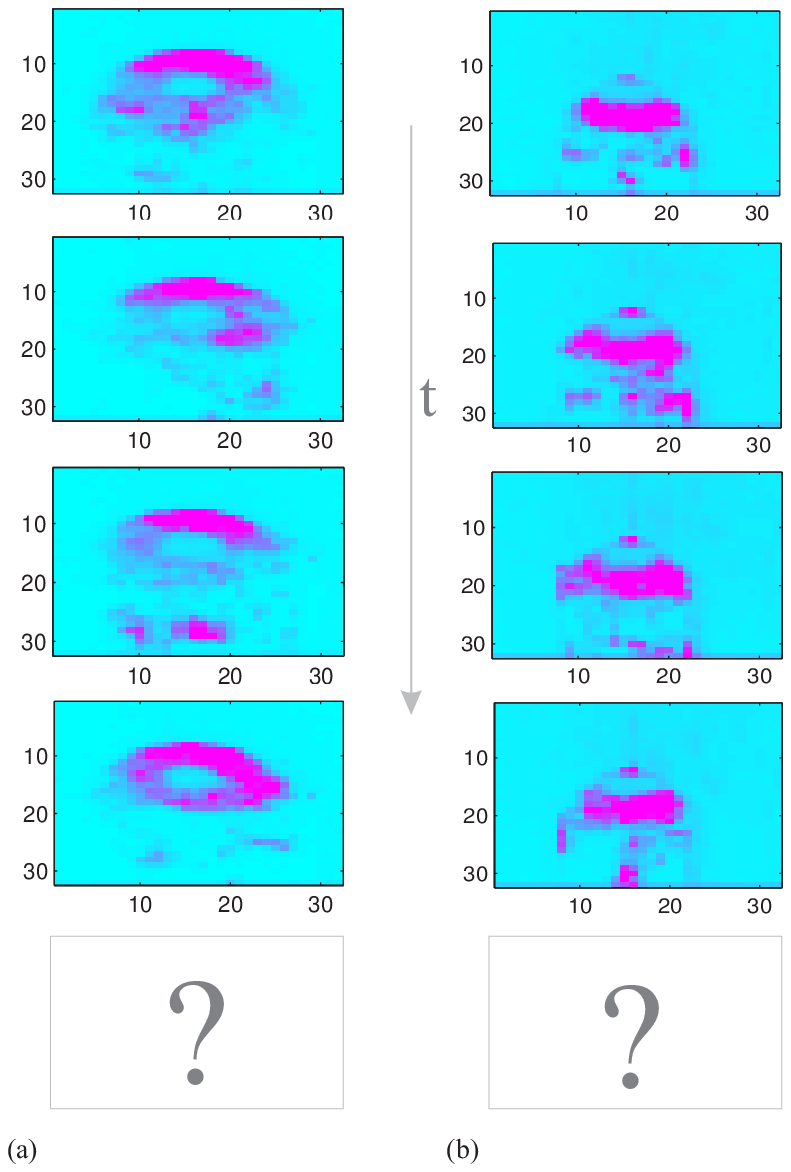}
\caption{Time series of laser welding records for two different welding 
regimes: (a) a deep welding regime , and (b) a heat conduction welding regime. 
The next time-step images denoted by {\em ?} are unknown and must be predicted.}
\label{slikazaporedje}
\end{figure}

Characteristic dynamic properties of laser welding process in space and 
time can be experimentally obtained by recording the surface 
optical activity of the melt pool. With respect to the 
energy supplied to the material, various dynamic regimes of the welding process 
can be distinguished. 
In Fig.\,\ref{slikazaporedje} visual records of two different welding regimes 
are shown, 
a deep welding regime (a) and a heat conduction welding regime (b). 
In the following discussion, only the deep welding regime is considered.

Dynamics of the welding regime are here represented by a record of 1000 images 
of size 32 $\times$ 32 points in space with sampling time 1/220\,s. This 
experimental record forms a three-dimensional field of light 
intensity ${\bf \varphi}$(${\bf s}={\bf r},t$) in two-dimensional 
space $\{(r_{x,i},r_{y,i}); i=1,..32,j=1..32\}$ and time $\{t_k;k=1..1000\}$. 
Due to local energy supply, the field is non-homogeneous in space.  Consequently, 
we model its evolution locally at each spatial point separately. 
A model of field evolution, {\em i.e.}, the {\em learning sample} 
is formed from the first 800 images.  We then predict the time evolution of 
the field and compare it with the next 200 images, which represent the 
{\em testing sample}. Based on the quality $Q$ of these predicted 
images, we optimize our prediction procedure, {\em i.e.}, and define 
the structure of the surroundings ${\cal S}$, and the optimal number 
of joint sample 
pairs $N$ and parameter $\sigma$.

\subsection{Optimal value of parameter $\sigma$}

\begin{figure}
\centering
\includegraphics[width=2.8in]{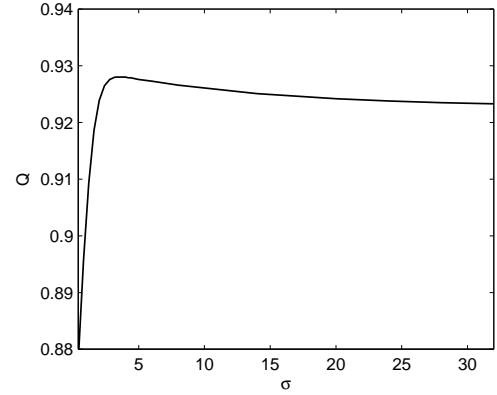}
\caption{Dependence of prediction quality $Q$ on the value of parameter $\sigma$ 
from {\em conditional average estimator} ${\bf \hat{y}}$. Surrounding set ${\cal S}$ is taken to be only one 
point with the same spatial position and a neighboring position in time.
Number of learning images in the learning field 
is set to 800.}
\label{QodSigma}
\end{figure}

Parameter $\sigma$ in the conditional average estimator ${\bf \hat{y}}$ 
(Eq.\,(\ref{CE})) was to this point  left undetermined. 
However, as shown in Fig.\,\ref{QodSigma}, obtained for the deep welding regime, 
the quality of prediction depends on the  value of $\sigma$. The learning sample 
consisted of 800 images and the surrounding of predicted field 
distribution in point ${\bf s} = ({\bf r},t)$ was taken to be 
just one neighboring  point with the same space coordinate 
and the time coordinate being one step behind ${\bf s}' = ({\bf r},t_{-1})$. 
Based on the learning sample, 
ten images from the testing interval have been predicted and compared with 
the corresponding images from the testing field. The value $Q$ shown in 
Fig.\,\ref{QodSigma} is the average quality of these ten images.

As Fig.\,\ref{QodSigma} shows, the quality exhibits a strong $\sigma$ dependence at the beginning of  the interval, and reaches 
its largest value for $\sigma $ approximately equal to 4. For larger values, it becomes a weakly 
decreasing function of $\sigma$. Since the optimal quality is reached for $\sigma \approx 4$,  this value is used in our further calculations.

As shown in Eq.\,(\ref{CE}), prediction of the field distribution at a given point is determined on the basis 
of similarity between the field distribution surrounding this point and the field distribution in  
surrounding points taken from the learning field. If we keep in mind that the parameter $\sigma$ defines the 
width of the Gaussian kernel function $\psi$ (see comment to Eq.\,\ref{PDF}), we can conclude, that for very small 
$\sigma$, only those joint sample pairs from the learning field which have a field distribution in the surrounding points (${\bf x_i}$) very similar to the field distribution in 
the surroundings of the point to be predicted (${\bf x}$) contribute to the predicted value of the field. 
On the other hand, for larger $\sigma$ those joint 
sample pairs with larger difference ${\bf x}-{\bf x_i}$ also contribute to the prediction of ${\bf \hat{y}}$. In the limit 
of large $\sigma$, almost all joint sample pairs contribute equally to ${\bf \hat{y}}$.

\subsection{Optimal number of joint sample pairs}

If the prediction of welding pool images is to be part of a laser welding control system, 
the prediction operation has to be performed in the shortest time interval possible. Since the number 
of operations needed to predict a field distribution in a given point increases linearly with the number 
of joint sample pairs (see Eq.\,\ref{CE}), it is necessary to find the smallest number of joint sample pairs 
which is still able to give predicted images of good quality. 

\begin{figure}
\centering
\includegraphics[width=2.8in]{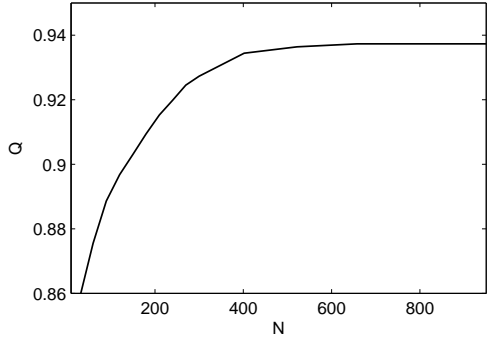}
\caption{Dependence of prediction quality $Q$ on number of images defining the learning field. 
The surrounding set ${\cal S}$ is taken to be only one point with the same spatial position 
and a neighboring position in time.
Parameter $\sigma$ is set to 4.}
\label{QodN}
\end{figure}

In Fig.\,\ref{QodN} we show the dependence of the quality of prediction on the number of images $N$ defining the learning field. 
As in the case of Fig.\,\ref{QodSigma}, the surrounding of the predicted field distribution in point ${\bf s} = ({\bf r},t)$ was taken 
to be just one neighboring point with the same space coordinate and the time coordinate being one step behind ${\bf s}' = ({\bf r},t_{-1})$. 
Parameter $\sigma$ is set to 4. Again $Q$ is taken to be the average quality of ten predicted images, which 
were compared with the corresponding images from the testing field.

As one can see from Fig.\,\ref{QodN}, $Q$ increases rather strongly with small values of $N$ ($N< 400$), while for $N>600$ an increase in $N$ does 
not result in a significant improvement of prediction quality. Therefore, in further calculations we apply $N=600$.

\subsection{Choosing the surrounding ${\cal S}$}

Our next goal is to find an optimal structure of RBFNN which yields the best quality of prediction in the shortest time interval. 
The structure of surrounding set ${\cal S}$ plays an important role in this optimization process since each additional point in 
the surrounding increases the dimensionality of vectors ${\bf x}_i$ and therefore
the time needed to predict the 
field distribution in a given point. Our task is to find the smallest surrounding of point ${\bf s}$, which results in  high prediction quality.

\begin{figure}
\centering
\includegraphics[width=3.2in]{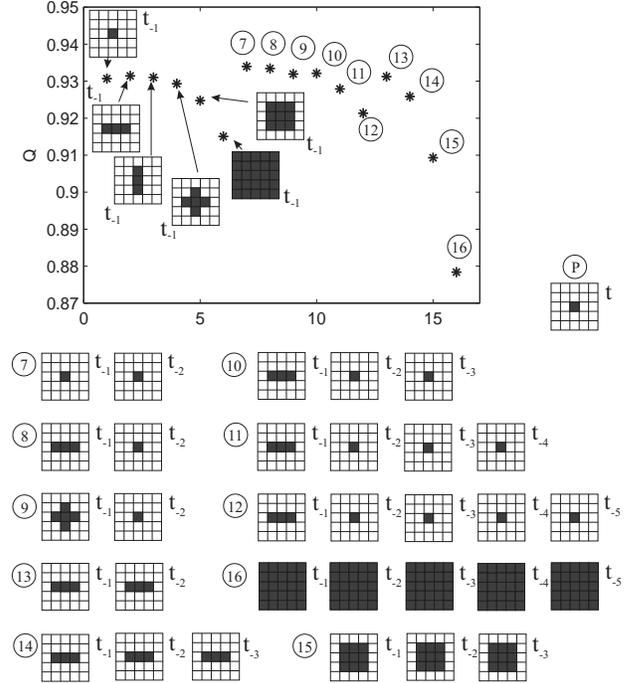}
\caption{Dependence of prediction quality $Q$ (*) on the structure of the surrounding set 
${\cal S}$. Parameters are $N=600$ and $\sigma=4$.
The netlike patterns (1-16) describe the position of surrounding points, while the 
netlike pattern denoted by P describes the position of prediction point ${\bf s}$. 
In the netlike patterns, only those time planes which contain 
points from ${\cal S}$ are plotted.
}
\label{Qods}
\end{figure}

In Fig.\,\ref{Qods} the prediction quality is presented for various selections of surrounding set ${\cal S}$. Parameters $N$ and  $\sigma$ are 600 and 4, respectively. 
As before, $Q$ represents the average quality of ten predicted images which were 
compared with the corresponding images from the testing field.  
All the member points of the first six surrounding sets in the diagram lie in the plane $t_{-1}$. Member points of other surrounding sets 
lie in several planes. For each of these, only those planes containing the member points are plotted.. 

As can be seen in  Fig.\,\ref{Qods}, the smallest surrounding sets give the 
best quality of prediction - see sets Nr. 1-3 and 7-10. 
If more points belonging to the same time-plane are added to ${\cal S}$, 
prediction quality is decreased- compare, for example sets Nr. 1 and Nr. 6 or Nr.\,14 and Nr.\,15. 
In contrast, surrounding sets containing points from two planes, $t_{-1}$ and
$t_{-2}$, give a slightly better $Q$ than sets containing only points from 
$t_{-1}$ - compare for example sets Nr.\,1 and Nr.\,7. 
However, an addition of multiple time-planes reduces the quality (see set Nr. 12).

As the best quality is obtained for set Nr.\,7, this surrounding set is 
considered optimal  in further calculations.
We would like to stress, that in Fig.\,\ref{Qods} only those surrounding sets which seemed to have the potential 
to give the best quality were taken into account. The optimal structure of ${\cal S}$ was chosen on the basis of selected sets. To be sure that the chosen structure 
was really optimal, it is necessary to calculate the prediction quality of all the subsets containing all combinations of neighboring points. 
Since the number of points in our learning set is 
32$\times$32$\times$600, a calculation of $Q$ for all sets would become a computationally prohibitive task. 

\subsection{Optimal prediction of melt pool evolution}

\begin{figure}
\centering
\includegraphics[width=3.4in]{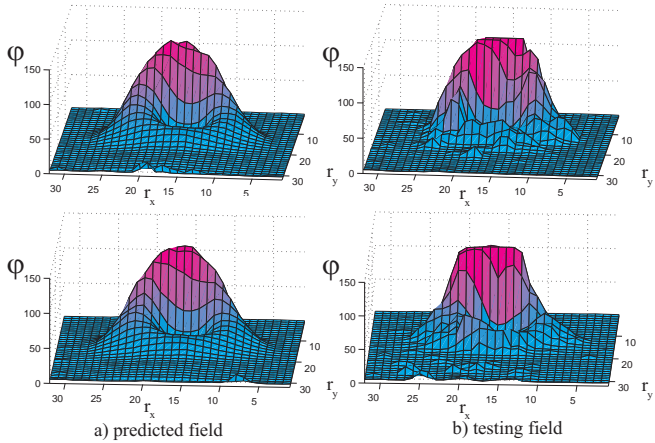}
\caption{Comparison of predicted melt pool images (a) with corresponding 
images from the testing field (b) for two randomly chosen testing records.
$\varphi$ stands for field, $r_x$ and $r_y$ denote spatial coordinates of the record. Parameters  are $N=600$ and $\sigma=4 $. 
The surrounding set ${\cal S}$ has only two member points, both having the same spatial position (${\bf r}$) as the predicted point ${\bf s}$=(${\bf r}$,t), 
but different neighboring positions in time, {\em i.e.},  $t_{-1}$ and $t_{-2}$. }
\label{slike}
\end{figure}

After determining the optimal parameters of our RBFNN model, we next show 
the discrepancy between the predicted images of the
laser welding melt pool  and the corresponding 
images from the testing field. In 
Fig.\,\ref{slike} we therefore present predicted laser welding 
images and corresponding images from the testing field for 
the optimal structure of RBFNN. Parameters are $N=600$, and $\sigma=4$, 
while the surrounding set  ${\cal S}$ has only two member points, 
both having the same spatial position as the predicted point, 
but neighboring positions in time. 
Since the quality of prediction is 0.93 (see Fig.\,\ref{Qods}), a very 
good similarity between the predicted and corresponding image from 
the testing field is expected. Comparison of 
predicted images and images from the testing field in Fig.\,\ref{slike} 
indeed exhibits a good resemblance. However, we would like to draw attention to surface 
smoothness. As can be seen, the predicted surface is smoother than the 
original surface. This can be easily understood if the origin of prediction 
of images in the conditional average estimator (Eq.\,\ref{CE}) is taken into 
account. Predicted ${\bf \hat{y}}$ is therefore a weighted average 
of all those ${\bf y}_i$, for which ${\bf x}_i$ is similar to ${\bf x}$. Consequently, the surface roughness is diminished due to conditional averaging.

\section{Conclusion}

Time evolution of multi-dimensional fields is usually obtained by solving a system of partial differential equations.
However, if the only source of information is a record of the field, a neural network can successfully replace differential
equations by extracting field evolution properties from the recorded data. Neural-network-like structures are also expected to 
be the working algorithm of living organisms' intelligence. In the same way as neural networks, living organisms predict 
the evolution of events in their surroundings solely on the basis of recorded data. It could be conjectured that 
this operation is probably performed 
by extracting simple evolution laws from recorded data. 

In this paper we show how to optimize a statistical modeling of 
a field generator performed by the normalized RBFNN, 
to efficiently learn spatiotemporal dynamics of  multi-dimensional fields. 
In our experimental approach, all information about process dynamics 
is contained in a measured space-time record of the characteristic variable.
To extract the model of field evolution from the corresponding discrete sample data, we 
employ a non-parametric approach, following a state-space reconstruction 
technique. The basis of state-space reconstruction is the formation of 
sample vectors which are composed of past and future field distributions.
We assume that the field distribution in a given spatiotemporal point 
${\bf s}$ is correlated with the field distribution in the spatiotemporal 
surroundings of this point, ${\cal S}$. 
The prediction of field distribution in ${\bf s}$ is then accomplished as 
a mapping relation between the field distribution in the surroundings 
${\cal S}$ and field distribution in ${\bf s}$.

Since the optimization of the state-space reconstruction technique also 
requires a quantitative measure of the prediction quality, we introduce the quality 
estimator $Q$, which incorporates the difference between the predicted field 
and the corresponding testing field. We consider as a proper set of model parameters 
those values at which the prediction quality achieves a maximum. This strategy is 
used here to find a proper value of parameter $\sigma$ and the structure of 
the surrounding ${\cal S}$ utilized in the prediction process. Generally, an 
estimation of the proper number of sample points must also consider the 
complexity of the experiments, which  is numerically demanding in a 
multidimensional case\,\cite{igextr}. Consequently, we also specify 
here  the proper number $N$ based upon the analysis of prediction quality.

We demonstrate the proposed method of modeling of the properties of the 
laser-heated melt pool. For this purpose, we employ 
non-parametric statistical modeling of 
field evolution on a spatiotemporal record of the melt pool of the 
laser welding process.
The major part of the field record is used for learning, while the minor 
part of the record serves for testing. We show how to 
construct the set of 
joint sample pairs containing past and future values of field distributions 
and pay special attention to the structure 
of these sample pairs. We also present the optimal structure of sample vectors, 
which gives the highest resemblance between 
predicted images and images from the testing field and has a 
small number of member points in order to make the prediction algorithm work quickly.


\section*{Acknowledgment}
This work was supported by the Ministry of Higher Education, Science and Technology of the Republic of Slovenia and EU-COST.


\begin{thebibliography}{1}

\bibitem{govekar1}
E.~Govekar, J.~Gradi\v sek, I.~Grabec, M.~Geisel, A.~Otto, M.~Geiger, 
"On characterization of CO$_2$ laser welding process by means of light emitted 
by plasma and images weld pool", in \emph{Third Int. 
Symp: Investigation of Non-linear Dynamics Effects in Production Systems}
(Cottbus, Germany), 2000.

\bibitem{govekar2}
E.~Govekar, J.~Gradi\v sek, I.~Grabec, M.~Geisel, A.~Otto, M.~Geiger, 
"Influence of feed rate on dynamics of laser welding process" in \emph{Second Int. 
Symp: Investigation of Non-linear Dynamics Effects in Production Systems}
(Aachen, Germany), 1999.



\bibitem{casdagli}
M.~Casdagli, S.~Eubank, "Nonlinear Modeling and Forcasting", Santa Fe Institute:
Addison-Wesley, 1992.

\bibitem{abarbanel}
H.~D.~I.~Abarbanel, R.~Brown, J.~J.~Sidorowich, L.~S.Tsmiring, "The analysis of 
observed chaotic data in physical systems",\emph{Rev. Mod. Phys.} Vol. 65, pp.1331-1392, 1993.

\bibitem{kosterlich}
E.~J.~Kosterlich, T.~Schreiber, "Noise reduction in chaotic time-series data: A survey to common methods
",\emph{Phys. Rev. E } Vol. 48, pp.1752-1763, 1993.

\bibitem{kantz}
H.~Kantz, T.~Schreiber, "Nonlinear Time Series Analysis", Cambridge University Press, 1997.

\bibitem{sigert}
S.~Sigert, R.~Friedrich, J.~Peinke, "Analysis of data sets of stochasitc systems",
\emph{Phys. Lett. A} Vol. 243, pp.275-289, 1998.

\bibitem{rubin}
D.~M.~Rubin, "Use of forecasting signatures to hlep destinguish periodicity, randomness, 
and chaos in ripples and other spatial patterns",
\emph{Chaos} Vol. 2, pp.525-535, 1992.

\bibitem{grabecmandelj}
I.~Grabec, S.~Mandelj, "Continuation of chaotic fields by RBFNN", in 
\emph{Bilological and Artificial Computation: From Neuroscience to Technology: Proc.},
eds. J.~Mira, R. Moreno-Diaz, J. Cebestany, Lecture Notes in Computer Science 
(Springer-Verlag, Berlin), Vol. 1240, pp.597-606, 1997.

\bibitem{orstavik}
S.~\O rstavik, J.~Stark, "Reconstruction and cross-prediction in coupled map lattices 
using spatiotemporal embedding techniques",
\emph{Phys. Lett. A} Vol. 247, pp.145-160, 1998.

\bibitem{parlitz}
U.~Parlitz, C.~Merkwirth, "Prediction of spatiotemporal time series based on 
reconstructed local states",
\emph{Phys. Rev. Lett.} Vol. 84, pp.1890-1893, 2000.


\bibitem{mandelj}
S.~Mandelj, I.~Grabec and E.~Govekar, "Statistical modeling of stochastic surface profiles",  
\emph{CIRP-J. Manuf. Syst}, Vol. 30, pp 281-287, 2000.

\bibitem{mandeljclanek}
S.~Mandelj, I.~Grabec and E.~Govekar, "Nonparametric statistical modeling of 
spatiotemporal dynamics based on recorded data",  \emph{Int. Jour. Bifur. Chaos}, 
Vol. 14, No. 6, pp 2011-2025, 2004.

\bibitem{duda3}
R.~O.~Duda, P.~E.~Hart, "Pattern Clasification and Scene analysis", J. Wiley and Sons, New York, Chap. 4. 1997.

\bibitem{vos98}
H.~U.~Voss, M.~J.~B\" unner, M.~Abel, "Identification of continuous, spatiotemporal 
systems", 
\emph{Phys. Rev. E} Vol. 57, pp.2820-2823, 1998.

\bibitem{bar}
M.~B\" ar, R.~Hegger, H.Kantz, "Fitting partial differential equations to space-time 
dynamics", 
\emph{Phys. Rev. E} Vol. 59, pp.337-342, 1999.

\bibitem{vos99}
H.~U.~Voss, P.~Kolodner, M.~Abel, J.~Kurths, "Amplitude equations from spatiotemporal 
binary-fluid convection data", 
\emph{Phys. Rev. Lett} Vol. 83, pp.3422-3425, 1999.

\bibitem{igextr} I.~Grabec, "Extraction of Physical Laws from Joint Experimental Data," \emph{Eur. Phys. J. B}, vol. 48, pp. 279-289, 2005.

\end{thebibliography}
\end{document}